\documentclass[12pt]{article}
\usepackage{epsfig}
\usepackage{psfrag}
\def\beq{\begin{equation}}
\def\eeq#1{\label{#1}\end{equation}}
\def\eeqn{\end{equation}}

\def\beqa{\begin{eqnarray}}
\def\eeqa#1{\label{#1}\end{eqnarray}}
\def\eeqan{\end{eqnarray}}

\let\bar=\overbar

\def\Dslash{\not{\hbox{\kern-4pt $D$}}}
\def\dslash{\not{\hbox{\kern-2pt $\del$}}}

\def\msb{{\bar{\ssstyle M \kern -1pt S}}}

\def\Title#1{\begin{center} {\Large {\bf #1} } \end{center}}

\begin{document}

\Title{D Physics}

\begin{center}{\large \bf Contribution to the proceedings of HQL06,\\
Munich, October 16th-20th 2006}\end{center}

\bigskip\bigskip


\begin{raggedright}  

{\it Svjetlana Fajfer\footnote{talk given by S. Fajfer}, Jernej Kamenik and Sa\v sa Prelov\v sek\index{
Fajfer, S}\\
Department of Physics, University of Ljubljana\\
and J. Stefan Institute\\
1000 Ljubljana, Slovenia}
\bigskip\bigskip
\end{raggedright}

\centerline{\bf Abstract}

\noindent
Recently a lot of new experimental results on open charm hadrons 
 have appeared. 
In particular many D meson resonances have been discovered.
 We discuss
strong decays of positive and negative parity charmed mesons
within heavy  meson chiral perturbation theory and study the impact
of excited charm states on the determination of the effective meson
couplings \cite{SF-JK-ch}.
Motivated by recent experimental results we also reconsider semileptonic 
$D \to P l \nu_l$ and $D \to V l \nu_l$ decays within a model which combines heavy
quark symmetry and properties of the chiral Lagrangian. 
Using limits of
soft collinear effective theory and heavy quark effective theory we
parametrize the semileptonic form factors. We include excited charm meson
states in our Lagrangians and determine their impact on the charm meson
semileptonic form factors.
In some scenarios of new physics an up-like heavy quark appears, which
induces FCNC at tree level for the $c \to u Z$ transitions. 
We investigate
FCNC effects in D rare decays in particular the $c \to u l^+ l^-$ transition
which might occur in $D^+ \to \pi^+  l^+ l^-$ and $D^0 \to \rho^0 l^+ l^-$.

\section{Strong decays of positive and negative parity charmed mesons}

The strong and electromagnetic transitions of  positive and negative parity charmed mesons  have 
already been studied within a variety of approaches  (see references [8] - 
[23] given in \cite{SF-JK-ch}). In ref.~\cite{Stewart} the chiral loop corrections to the $D^* \to D \pi$ and $D* \to D \gamma$ decays were calculated 
and a numerical extraction of the one-loop bare couplings was first performed. Since this calculation preceded the discovery of even-parity meson states,
it did not involve loop contributions containing the even-parity meson states. 
The ratios of the radiative and strong decay widths, and the 
isospin violating decay $D_s^* \to D_s \pi^0$ were used to extract the relevant couplings. 
However, since that time, the experimental situation has improved and therefore we consider  the chiral loop contributions to the strong decays of the even and odd parity charmed meson states using HH$\chi$PT. 
In our calculation we consider the strong decay modes $D^{*+}$, $D^{*0}$, $D^{*+}_0$  $D^{*0}_0$ and 
$D^{'0}_1$ (given in Table~1 of~\cite{SF-JK-ch}) 
\par
The existing data on the decay widths enable us to constrain the leading order parameters: the $D^* D\pi$ coupling $g$, $D^{*}_0 D \pi$ coupling $h$, and the coupling $\tilde g$ which enters in the interaction of even parity charmed mesons and the light pseudo-Goldstone bosons. Although the coupling 
$\tilde g$ is not yet  experimentally constrained, it moderatelly affects the decay amplitudes which we investigate. 
\par
\begin{figure}[!t]
\begin{center}
\epsfig{file=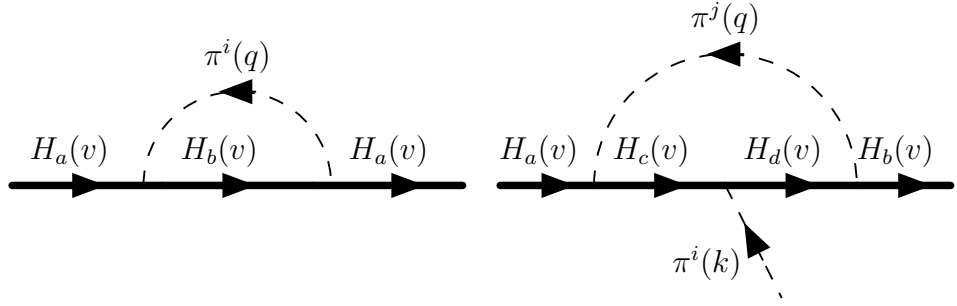,width=5in}
\end{center}
\caption{\label{fig_1}Sunrise (left) and sunrise road (right) topology diagrams}
\end{figure}

Due to the divergences coming from the chiral loops one needs to include the appropriate counterterms. 
Therefore we construct a full operator basis of the relevant counterterms and include it into our 
effective theory Lagrangian. 
The details of the Heavy hadron chiral perturbation theory  HH$\chi$PT we use is given in \cite{SF-JK-ch}. First we determine wavefunction renormalization of the relevant heavy meson fields considering the effects of the chiral loops given by the left diagram in Fig.~\ref{fig_1}.
\psfrag{pijl}[bc]{$\pi^i(q)$}
\psfrag{Ha}[bc]{$H_a(v)$}
\psfrag{Hb}[bc]{$H_b(v)$}
\psfrag{pi}[bc]{$\pi^i(k)$}
\psfrag{Hc}[bc]{$H_c(v)$}
\psfrag{Hd}[bc]{$H_d(v)$}
\psfrag{pj}[bc]{$\pi^j(q)$}
Then we calculate loop corrections for the $PP^*\pi$, $P_0 P^*_1 \pi$ and $P_0 P\pi$ vertexes. At zeroth order in $1/m_Q$ expansion these are identical to the $P^*P^*\pi$, $P^*_1 P^*_1 \pi$ and $P^*_1P^*\pi$ couplings respectively due to heavy quark spin symmetry (right diagram in Fig.~\ref{fig_1}).
\par
Using known experimental values for the decay widths of $D^{+*}$, $D^{+*}_0$, $D^{0*}_0$ and $D^{'}_1$, and the upper bound on the width of $D^{0*}$ one can extract the values for the bare couplings $g$, $h$ and $\tilde g$ from a fit to the data. The decay rates are namely given by
\begin{equation}
\Gamma(P_a^{*}\to \pi^i P_b) = \frac{|g^{\mathrm{eff.}}_{P^{*}_a P_b\pi^i}|^2}{6 \pi f^2} |\vec k_{\pi^i}|^3,
\end{equation}
and a similar expression (up to polarization averaging phase space factors) for $\Gamma(P_0\to \pi P)$ and $\Gamma(P^*_1\to \pi P^*)$ with $g$ coupling replaced by $h$ and $|\vec k_{\pi}|^3$ replaced by $E^2_{\pi}|\vec k_{\pi}|$. Here $\vec k_{\pi}$ is the three-momentum vector of the outgoing pion and $E_{\pi}$ its energy. The renormalization condition for the couplings can be written as
\begin{equation}
g^{\mathrm{eff.}}_{P_a^*P_b\pi^i} = g \frac{\sqrt{Z_{2P_a}}\sqrt{Z_{2P_b^*}}\sqrt{Z_{2\pi^i}}}{\sqrt{Z_{1P_aP_b^*\pi^i}}} = g Z^g_{P_a^*P_b\pi^i}
\label{eq_g_renorm}
\end{equation}
with similar expressions for the $h$ and $\tilde g$ couplings. 
\par
We perform a fit with a renormalization scale set to $\mu\simeq1~\mathrm{GeV}$~\cite{Stewart} and we neglect counterterm contributions altogether. Our choice of the renormalization scale in dimensional regularization is arbitrary and depends on the renormalization scheme. Therefore any quantitative estimate made with such a procedure cannot be considered meaningful without also thoroughly investigating counterterm, quark mass and scale 
dependencies. We perform a Monte-Carlo randomized least-squares fit for all the three couplings in the prescribed 
regions \cite{SF-JK-ch} using the experimental values for the decay rates to compute $\chi^2$ and using values from PDG \cite{PDG} for the masses of final state heavy and light mesons. In the case of excited $D^*_0$ and $D'_1$ mesons, we also assume saturation of the measured decay widths with the strong decay channels to 
ground state charmed mesons and pions ($D^*_0 \to D \pi$ and $D'_1\to D^* \pi$). 
\par
Due to the rather large mass splitting between positive in negative parity states $\Delta_{SH}$, we find that the
perturbative expansion holds for scales below $\mu \leq 1~\mathrm{GeV}$, while these new strongly scale dependent corrections become large at higher renormalization scales. 
From the data for the four decay widths we obtain the best-fitted values for the bare couplings, which we summarize in Table ~\ref{table_summary}. 
\begin{table}[!t]
\begin{center}
\begin{tabular}{lccc}
\hline \hline
 Calculation scheme & $g$ & $|h|$ & $\tilde g$ \\
 \hline
 Leading order & $0.61$  & $0.52$ & $-0.15$\\
 One-loop without positive parity states & $0.53$ &  &  \\
 One-loop with positive parity states  & $0.66$ & $0.47$ & $-0.06$ \\
\hline \hline
\end{tabular} 
\caption{Summary of our results for the effective couplings as explained in the text. The listed best-fit values for the one-loop calculated bare couplings were obtained by neglecting counterterms' contributions at the 
regularization scale $\mu\simeq 1~\mathrm{GeV}$.}
\label{table_summary}
\end{center}
\end{table} 
We are able to determine all the three couplings since the contributions proportional to the coupling $\tilde g$ 
appear  indirectly, through the loop corrections.
\par
Since we consider decay modes with the  pion in the final state, we do not expect 
sizable  contribution of the counterterms. Namely, the counterterms which appear in our study 
are  proportional to the light quark masses, and not to the strange quark mass~\cite{Stewart}. The effects of counterterm contributions in the decay modes we analyze, are nevertheless estimated by making the random distribution of the relevant counterterm couplings (see \cite{SF-JK-ch}). The counterterm contributions of order $\mathcal O (1)$ can 
spread the best fitted values of $ g$, $|h|$ by roughly $15\%$ and $\tilde g$ by as much as $60\%$. Similarly, up to $20\%$ shifts in the renormalization scale modify the fitted values for the $g$ and $|h|$ by less than $10\%$ while $\tilde g$ may even change sign at high renormalization scales. Combined with the estimated $20\%$ uncertainty due to discrepancies in the measured excited heavy meson masses, we consider these are the dominant sources of error in our determination of the couplings. 
One should keep in mind however that without better experimental data and/or lattice QCD inputs, the phenomenology of strong decays of charmed mesons presented above ultimately cannot be considered reliable at this stage.
\par
A full calculation of the strong decay couplings should also contain, in addition to the calculated contributions, the relevant $1/m_H$ corrections as discussed in ref~\cite{mehen}. There, the next to leading terms ($1/m_H$) were included in the study of charm meson mass spectrum.  
Due to the very large number of unknown couplings the combination of $1/m_H$ and chiral 
corrections does not seem to be possible for the decay modes we consider. In addition, 
recent studies of the lattice QCD groups~\cite{Abada,McNeile}
indicate that the $1/m_H$ corrections do not contribute significantly to 
their determined values of the strong couplings, 
and we therefore assume the same to be true in our calculations of chiral corrections. 
\par
Due to computational problems associated with the chiral limit, lattice QCD studies perform calculations at large light quark masses and then employ a chiral extrapolation $m_\pi \to 0$ of their results.
Our analysis of such chiral extrapolation of the coupling $g$ shows that 
the full loop contributions of excited charmed mesons give 
sizable effects in modifying the slope and curvature in the limit $m_\pi \to 0$. We argue that this is due to the inclusion of hard pion momentum scales inside chiral loop integrals containing the large mass splitting between charmed mesons of opposite parity $\Delta_{SH}$ which does not vanish in the chiral limit. If we instead impose physically motivated approximations for these contributions - we expand them in terms of $1/\Delta_{SH}$ - the effects reduce mainly to the changes in the determined values of the bare couplings, used in the extrapolation, with explicit $h$ contributions shrinking to the order of 5\%. Consequently one can infer on the good convergence of the $1/\Delta_{SH}$ expansion.
\par
As a  summary of our results, we point out  that chiral loop corrections in strong charm meson decays 
can be kept under control, but they give important contributions and are relevant for 
 the precise extraction of the strong coupling constants $g$, $h$ and $\tilde g$.

\section{Charm meson resonances in $D$ semileptonic decays}

The knowledge of the form factors which describe the weak $heavy \to
light$ semileptonic transitions is very important for the accurate
determination of the CKM parameters from the experimentally measured
exclusive decay rates. Usually, the attention has been devoted to 
$B$ decays and the determination of the phase of the $V_{ub}$ CKM matrix element.
At the same time in the charm sector, the most accurate determination of the size of $V_{cs}$ and $V_{cd}$ matrix elements is not from a direct measurement, mainly due to theoretical uncertainties  in the calculations of the relevant form factors' shapes.
\par 
Recently, there have been new interesting results on $D$-meson
semileptonic decays.  The CLEO and FOCUS collaborations have studied semileptonic
decays $D^0\rightarrow \pi^- \ell^+ \nu$ and 
$D^0\rightarrow K^- \ell^+ \nu$~\cite{Huang04,Link04}. 
Their data provide new information on the $D^0\rightarrow \pi^- \ell^+ \nu$ and $D^0\rightarrow K^- \ell^+ \nu$ form factors. Usually in $D$ semileptonic decays a simple pole parametrization was used in the past. The results of Refs.~\cite{Huang04,Link04} for the single pole parameters required by the fit of their data, however, suggest pole masses, which are inconsistent with the physical masses of the lowest lying charm meson resonances. In their analyses they also utilized a modified pole fit as suggested in~
\cite{Becirevic99} and their results indeed suggest the existence of
contributions beyond the lowest lying charm meson resonances~\cite{Huang04}.
\par
In addition to these results new experimental studies of 
charm meson resonances have provided a lot of new information on the charm sector
~\cite{Aubert03,Vaandering04,Besson03,Evdokimov04} 
which we can now apply to $D$ and $D_s$ semileptonic decays.
\par
The purpose of our studies~\cite{Fajfer04,Fajfer05,Fajfer06} is to accommodate  
contributions of the newly
discovered and theoretically predicted charm mesons in form factors which 
are parametrized
using constraints coming from heavy quark effective theory (HQET) limit for the region of 
$q_{max}^2$ and in the $q^2 \simeq 0$ region using results of 
soft collinear effective theory (SCET).
We restrain our discussion to the leading chiral and $1/m_H$ terms in the expansion.

\par

The standard decomposition of the current matrix elements 
relevant to semileptonic decays between a heavy pseudoscalar meson state 
$|H(p_H)\rangle$ with momentum $p_H^{\nu}$ and a light pseudoscalar meson state $| P (p_P) \rangle$ with momentum $p_P^{\mu}$ is in terms of two scalar functions of the exchanged momentum squared $q^2 = (p_H-p_P)^2$ -- the form factors $F_+(q^2)$ and $F_0(q^2)$. Here $F_+$ denotes the vector form factor and it is dominated by vector meson resonances, while $F_0$ denotes
the scalar form factor and is expected to be dominated by scalar meson
resonance exchange~\cite{Marshak69,Wirbel85}. 
In order that the matrix elements are finite at $q^2=0$, the form factors must
also satisfy the relation $F_+(0)=F_0(0)$.
\par
The transition of $|H(p_H)\rangle$ to light vector meson 
$|V(p_V,\epsilon_V)\rangle$ with momentum 
$p_V^{\nu}$ and polarization vector $\epsilon_V^{\nu}$ is similarly parameterized in terms of four form factors $V$, $A_0$, $A_1$ and $A_2$, again functions of the exchanged momentum squared $q^2 = (p_H-p_V)^2$. Here $V$ denotes the vector form factor and is expected to be dominated by vector meson resonance exchange, the axial $A_1$ and $A_2$ form factors are expected to be dominated by axial resonances, while $A_0$ denotes the 
pseudoscalar form factor and is expected to be dominated by pseudoscalar 
meson resonance exchange~\cite{Wirbel85}.
As in previous case in order that the matrix elements are finite at $q^2=0$, the form factors must also satisfy the well known relation $A_0(0)+A_1(0)(m_H+m_V)/2m_V-A_2(0)(m_H-m_V)/2m_V=0$.
\par
Next we follow the analysis of Ref.~\cite{Becirevic99}, where the $F_+$
form factor in $H\to P$ transitions is given as a sum of two pole contributions, while the
$F_0$ form factor is written as a single pole. This 
parametrization includes all known properties of form factors at large $m_H$.  Using a relation which connects the  form factors within large energy release approach~
\cite{Charles98} the authors in Ref.~\cite{Becirevic99} propose the following form factor 
parametrization
\begin{equation}
F_+(q^2)=\frac{F(0)}{(1-x)(1-a x)}, \qquad F_0(q^2)=\frac{F(0)}{1-b x},\label{f_+_dipole}
\end{equation}
where $x=q^2/m_{H^*}^2$. 
\par
Utilizing the same approach we propose a general parametrization of the heavy to light 
vector form factors, which also takes into account all the known scaling and resonance 
properties of the form factors \cite{Fajfer05,Fajfer06}
As already mentioned, there exist the well known HQET scaling laws in 
the limit of zero recoil~\cite{Isgur90} while in the SCET limit $q^2\to 0$  one obtains that all four $H \to V$ form factors can be related to only two universal SCET 
scaling functions~\cite{Charles98}.
\par
The starting point is the vector form factor $V$, which is dominated by the pole at $t=m_{H^*}^2$ when considering the part of the phase space that is close to the zero recoil. For the $heavy\to light$ transitions this situation is expected to be 
realized near the zero recoil where also the HQET scaling  applies. 
On the other hand, in the region of large recoils, SCET dictates the 
scaling described in \cite{Charles98}. In the full analogy with the 
discussion made in Refs. \cite{Becirevic99,Hill05}, the vector 
form factor consequently receives contributions from two poles and can be 
written as
\begin{equation}
V(q^2) = \frac{V(0)}{(1-x)(1-a x)},
\label{eq_v_ff}
\end{equation}
where $x=q^2/m_{H^*}^2$ ensures, that the form factor is 
dominated by the physical $H^*$ pole, while $a$ measures 
the contribution of higher states which are parametrized by another 
effective pole at $m_{\mathrm{eff}}^2=m_{H^*}^2/a$.  
\par
An interesting and useful feature one gets from the 
SCET is the relation between $V$ and $A_1$~\cite{Charles98,Ebert01,Burdman00, 
Hill04} at $q^2\approx 0$. When combined with our result~(\ref{eq_v_ff}), it imposes a single pole structure on $A_1$. We can thus continue in the same line of argument and write
\begin{equation}
A_1(q^2) =  \xi \frac{V(0)}{1-b' x}.
\label{eq_a1_ff}
\end{equation}
Here $\xi=m_H^2/(m_H+m_V)^2$ is the proportionality factor between $A_1$ and $V$ from the SCET relation, while $b'$ measures the contribution of resonant states with spin-parity assignment $1^+$ which are parametrized by the effective pole at $m_{H'^*_{\mathrm{eff}}}^2=m_{H^*}^2/b'$. It can be readily checked that also $A_1$, when parametrized in this way, satisfies all the scaling constraints. 
\par
Next we parametrize the $A_0$ form factor, which is completely 
independent of all the others so far as it is dominated by the pseudoscalar 
pole and is proportional to a different universal function in SCET. 
To satisfy both HQET and SCET scaling laws we parametrize it as 
\begin{equation}
A_0(q^2) = \frac{A_0(0)}{(1-y)(1-a' y)},
\label{eq_a0_ff}
\end{equation}
where $y = q^2/m_H^2$ ensures the physical $0^-$ pole dominance at small 
recoils and $a'$ again parametrizes the contribution of 
higher pseudoscalar states by an effective pole at 
$m_{H'_{\mathrm{eff}}}^2=m_{H}^2/a'$. The resemblance to $V$ is 
obvious and due to the same kind of analysis~\cite{Becirevic99} although 
the parameters appearing in the two form factors are completely unrelated. 
\par
Finally for the $A_2$ form factor, due to the pole behavior of the $A_1$ 
form factor on one hand and different HQET scaling at 
$q^2_{\mathrm{max}}$ on the other hand, we have to go beyond 
a simple pole formulation. Thus we impose
\begin{equation}
A_2(q^2) = \frac{A_2(0)}{(1-b' x)(1-b'' x)},
\label{eq_a2_ff}
\end{equation}
which again satisfies all constraints. Due to the relations between the form factors we only gain one parameter in this formulation, $b''$. This however causes the contribution of the $1^+$ resonances to be shared between the two effective poles in this form factor.
\par
At the end we have parametrized the four $H\to V$ vector form factors in 
terms of the six parameters $V(0)$, $A_0(0)$, $a$, $a'$, $b'$ and 
$b''$ ($A_2(0)$ is fixed by the kinematical constraint).

\par

In our heavy meson chiral theory (HM$\chi$T) calculations we use the leading order heavy meson chiral  Lagrangian in which we include additional charm meson resonances.
The details of this framework are given in \cite{Fajfer04}  and \cite{Fajfer05}. 
We first calculate values of the form factors in the small recoil region. 
The presence of charm meson resonances in our Lagrangian affects the values of
the form factors at $q^2_{\mathrm{max}}$ and induces saturation of the second 
poles in the parameterizations of the $F_+(q^2)$, $V(q^2)$ and $A_0(q^2)$ form 
factors by the next radial excitations of $D_{(s)}^*$ and $D_{(s)}$ mesons respectively. 
Although the $D$ mesons mat not be considered heavy enough, we employ these parameterizations with model matching conditions at $q^2_{\mathrm{max}}$.
Using HQET parameterization of the current matrix elements~\cite{Fajfer04,Fajfer05}, 
which is especially suitable for HM$\chi$T calculations of the form factors near zero 
recoil, we are able to extract consistently the contributions of individual resonances from our Lagrangian to the various $D\to P$ and $D\to V$ form factors. 
We use physical pole masses of excited state charmed mesons in the extrapolation, giving for 
the pole parameters $a=m_{H^{*}}^2/m_{H'^{*}}^2$, $a'=m_{H}^2/m_{H'}^2$, $b'=m_{H^*}^{2}/m_{H_{A}}^2$.
Although in the general parameterization of the form factors the extra poles in 
$F_+$, $V$ and $A_{0,1,2}$ parametrized all the neglected higher resonances beyond the ground state heavy meson spin doublets $(0^-,1^-)$, we are here 
saturating those by a single nearest resonance.
The single pole $q^2$ behavior of the $A_1(q^2)$ form factor is explained 
by the presence of a single $1^+$ state relevant to each decay, while in 
$A_2(q^2)$ in addition to these states one might also account for their next 
radial excitations. However, due to the lack of data on their presence we 
assume their masses being much higher than the first $1^+$ states and we 
neglect their effects, setting effectively $b''=0$.
\par
The values of the new model parameters appearing in $D \to P l \nu_l$ 
decay amplitudes~\cite{Fajfer04} 
are determined by fitting the model predictions to known experimental
values of branching ratios $\mathcal B (D^0\rightarrow K^- \ell^+
\nu)$, $\mathcal B (D^+\rightarrow \bar K^0 \ell^+ \nu)$, $\mathcal B
(D^0\rightarrow \pi^- \ell^+ \nu)$, $\mathcal B (D^+\rightarrow \pi^0
\ell^+ \nu)$, $\mathcal B (D^+_s\rightarrow \eta \ell^+ \nu)$ and
$\mathcal B (D^+_s\rightarrow \eta' \ell^+
\nu)$~\cite{PDG}. In our  calculations of decay widths we 
neglect the lepton mass, so the form factor $F_0$, which is
proportional to $q^{\mu}$, does not contribute. For the decay width we
then use the integral formula proposed in~\cite{Bajc95} with the flavor mixing parametrization of the weak current defined in~\cite{Fajfer04}.
\par
Similarly in the case of $D \to V l \nu_l$ transitions we have to 
fix additional model parameters~\cite{Fajfer05} and we again 
use known experimental values of branching
 ratios $\mathcal B (D_0\rightarrow K^{*-}\ell^+\nu)$, $\mathcal 
B (D_s^+ \rightarrow \Phi\ell^+\nu)$, $\mathcal B (D^+\rightarrow 
\rho^0\ell^+\nu)$, $\mathcal B (D^+\rightarrow K^{*0}\ell^+\nu)$, as well 
as partial decay width ratios $\Gamma_L/\Gamma_T (D^+\rightarrow K^{*0}
\ell^+\nu)$ and $\Gamma_+/\Gamma_- (D^+\rightarrow K^{*0}\ell^+\nu)$
~\cite{PDG}. We calculate the decay rates for polarized final light vector mesons using helicity 
amplitudes $H_{+,-,0}$ as in for example~\cite{Ball91}. By neglecting the lepton masses we again arrive at the integral expressions from~\cite{Bajc95} with the flavor mixing parametrization of the weak current defined in~\cite{Fajfer05}.

\par

\begin{figure}[!t]
\begin{tabular}{cc}
\epsfig{file=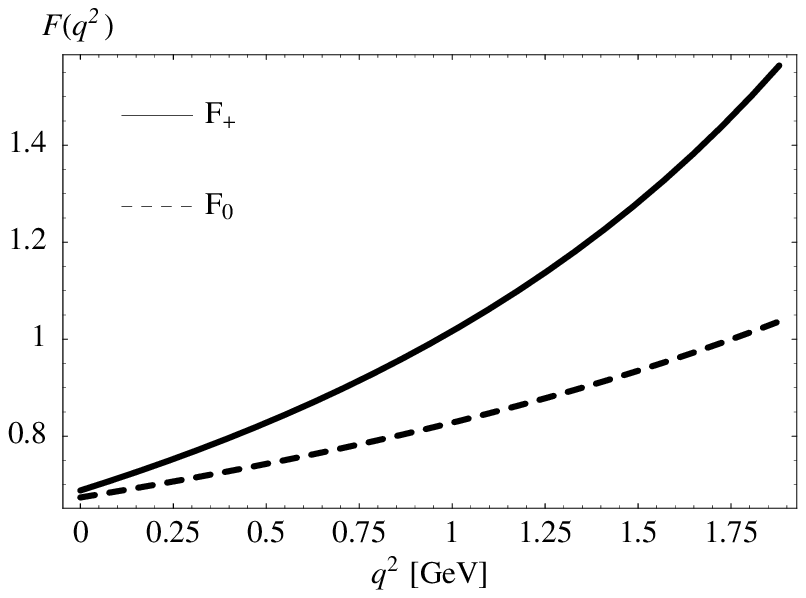,width=3.in} & 
\epsfig{file=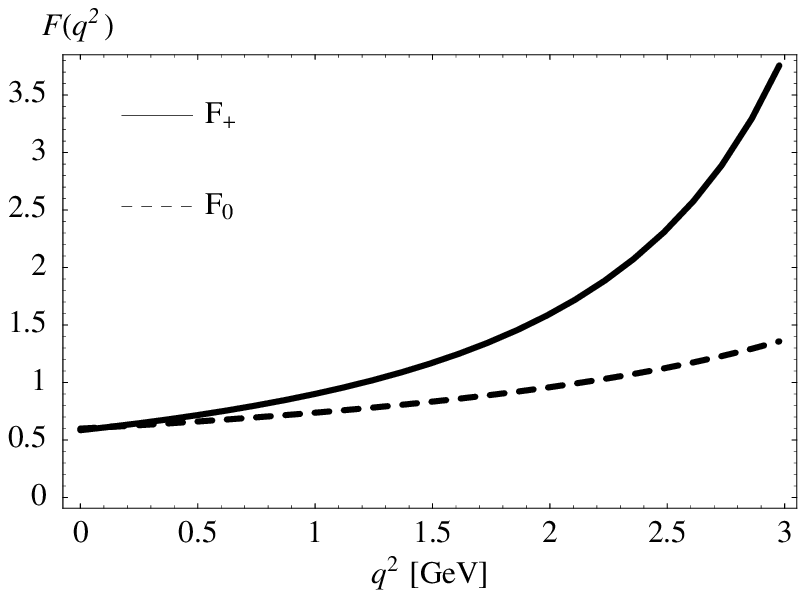,width=3.in} \\
\multicolumn{2}{c}{\epsfig{file=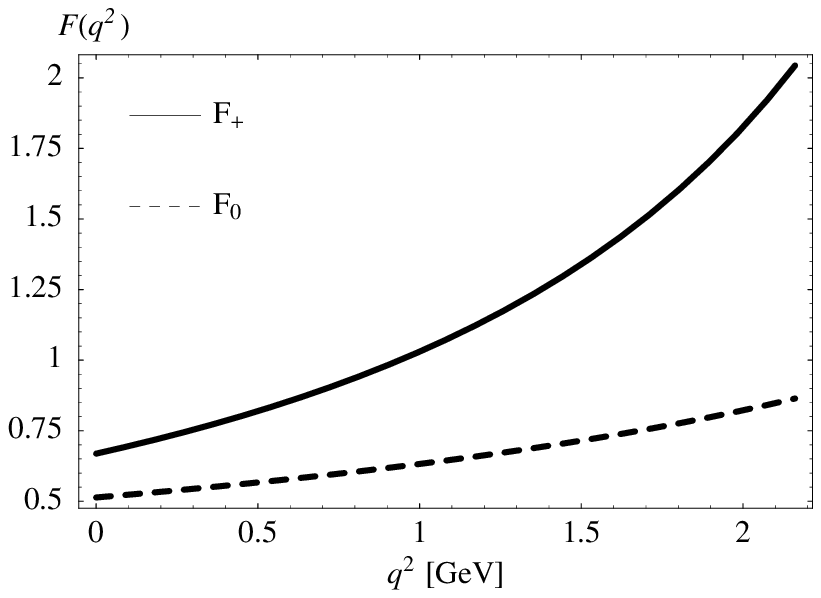,width=3.in}} 
\end{tabular}
\caption{\label{FplotDK}$q^2$ dependence of the $D^0 \rightarrow K^-$ (upper left), $D^0 \rightarrow \pi^-$ (upper right) and $D_s \rightarrow K^0$ (lower) transition form factors.}
\end{figure}

We first draw  the $q^2$ dependence of the  $F_+$ and $F_0$ form factors for the 
$D^0\rightarrow K^-$, $D^0\rightarrow \pi^-$ and $D_s\rightarrow K^0$ transitions. The results are depicted in Fig.~\ref{FplotDK}.
Our model results, when extrapolated with the double pole parameterization, 
agree well with previous theoretical~\cite{Melikhov00,Aubin04} and 
experimental~\cite{Huang04,Link04} studies whereas the single pole extrapolation does not give satisfactory results.
Note that without the scalar resonance, one only gets a soft pion contribution to the $F_0$ form factor. This gives for the $q^2$ dependence of $F_0$ a constant value for 
all transitions, which largely disagrees 
with lattice QCD results~\cite{Aubin04} as well as heavily violates 
known form factor relations.
\par
We also calculate the
branching ratios for all the relevant $D\rightarrow P$ semileptonic
decays and compare the predictions of our model with experimental data
from PDG. The results are summarized in Table~\ref{PP_results_table}. 
For comparison we also include the results for the rates obtained with
our approach for $F_+(q_{\mathrm{max}}^2)$  but using a 
single pole fit. 
\begin{table}[!t]
\begin{center}
\begin{tabular}{l|ccc}
\hline \hline
Decay & $\mathcal{B}$ (Mod. double pole) [\%] & $\mathcal{B}$ (Mod. single pole)
 [\%] & $\mathcal{B}$ (Exp. PDG) [\%] \\
\hline 	
	$D^0\to K^-$ & $3.4$ & $4.9$ & $3.43 \pm 0.14$ \\
	$D^0\to \pi^-$ & $0.27$ & $0.56$ & $0.36 \pm 0.06$ \\
	$D_s^+\to \eta$ & $1.7$ & $2.5$ & $2.5 \pm 0.7$ \\
	$D_s^+\to \eta'$ & $0.61$ & $0.74$ & $0.89 \pm 0.33$ \\
	$D^+\to \bar K^0$ & $9.4$ & $12.4$ & $6.8 \pm 0.8$ \\
	$D^+\to \pi^0$ & $0.33$ & $0.70$ & $0.31 \pm 0.15$ \\
	$D^+\to \eta$ & $0.10$ & $0.15$ & $<0.5$ \\
	$D^+\to \eta'$ & $0.016$ & $0.019$ & $<1.1$ \\
	$D_s^+\to K^0$ & $0.20$ & $0.32$ &  \\
\hline \hline
\end{tabular}
\end{center}
\caption{\label{PP_results_table} The branching ratios for the $D\rightarrow P$ semileptonic decays. Comparison of our model fit with experiment as explained in the text.}
\end{table}
It is very interesting that our model extrapolated with a double
pole gives branching ratios for $D \to P \ell \nu_{\ell}$ in rather good
agreement with experimental results for the already measured decay
rates. It is also obvious that the single pole fit 
gives the rates up to a factor of two larger than the experimental results. 
Only for decays to $\eta$ and $\eta'$ as given in  
Table~\ref{PP_results_table}, an agreement with experiment of the double pole 
version of the model is not better but worse than for the single pole case.
\par
\begin{figure}[!t]
\begin{tabular}{cc}
\epsfig{file=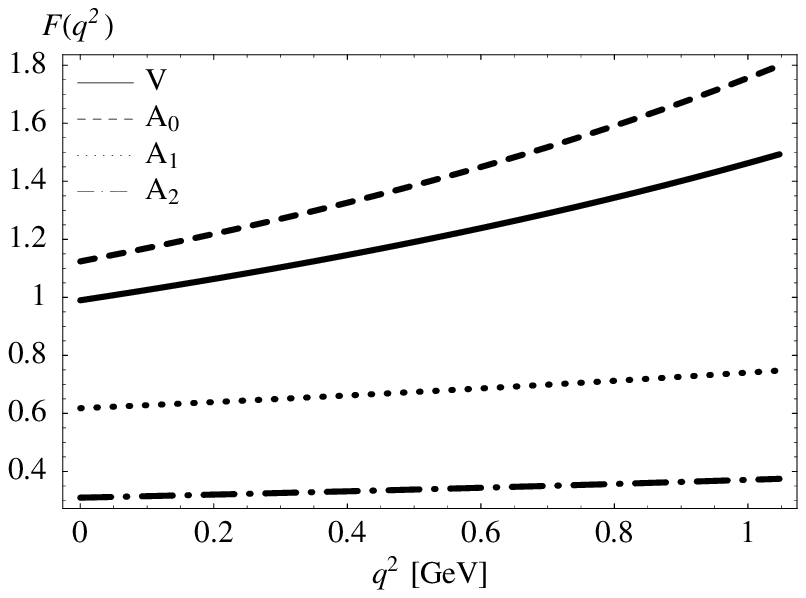,width=3.in} & 
\epsfig{file=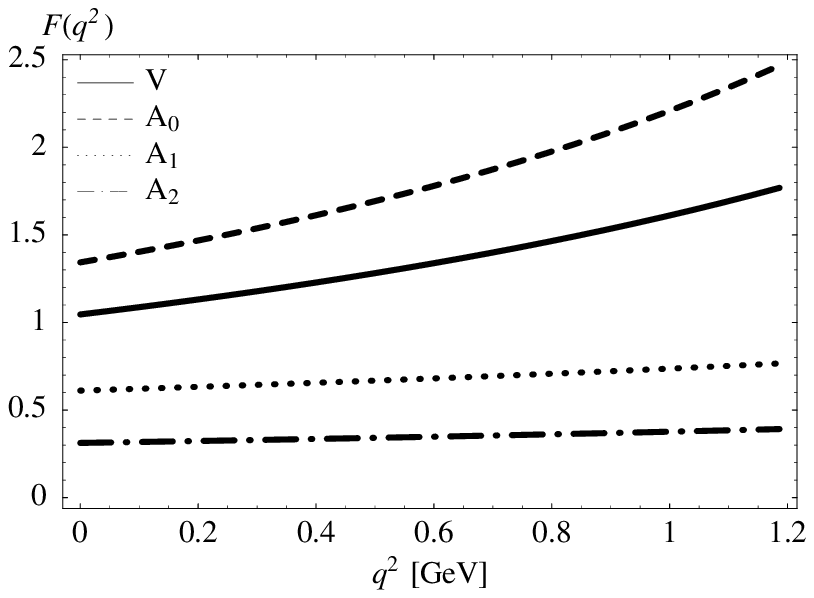,width=3.in} \\
\multicolumn{2}{c}{\epsfig{file=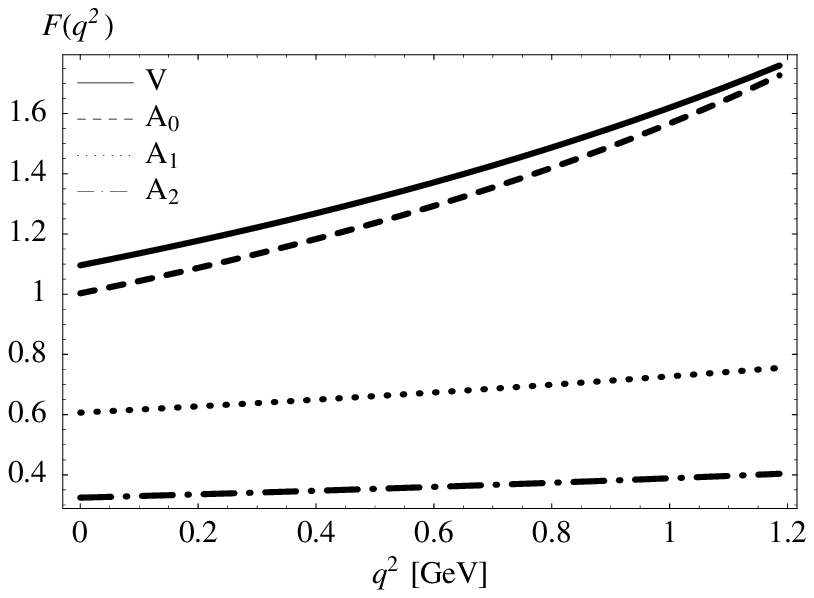,width=3.in}} 
\end{tabular}
\caption{\label{FplotDKs}$q^2$ dependence of the $D^0 \rightarrow K^{*-}$ (upper left), $D^0 \rightarrow \rho^{-}$ (upper right) and $D_s \rightarrow \phi$ (lower) transition form factors.}
\end{figure}

We next draw the $q^2$ dependence of all the form factors for 
the $D^0\to K^{-*}$, $D^0\to \rho^-$ and $D_s \to \phi$ transitions. The results are 
depicted in Fig.~\ref{FplotDKs}.
Our extrapolated results for the shapes of the $D\to V$ semileptonic form factors agree well with existing theoretical studies~\cite{Ball91,Melikhov00,Abada02,Ball93}, while currently no experimental determination of the form factors' shapes in these decays exists.  
\par
We complete our study by calculating branching ratios and partial decay width 
ratios also for all relevant $D \to V \ell \nu_{\ell}$ decays. 
They are listed in Table~\ref{table_results} together with known 
experimentally measured values.  
\begin{table}[!b]
\begin{center}
\begin{tabular}{l|cccc}
\hline \hline
Decay & $\mathcal{B}$ (Mod.) [\%] & $\mathcal{B}$ (Exp.) [\%] & 
$\Gamma_L/\Gamma_T$ (Mod.)&  $\Gamma_+/\Gamma_-$ (Mod.) \\
\hline 	
	$D_0\to K^*$ & $2.2$ & $2.15 \pm 0.35$~\cite{PDG} & $1.14$ & $0.22$ \\
	$D_0\to \rho$ & $0.20$ & $0.194\pm 0.039 \pm 0.013$~\cite{Blusk05} & $1.11$ &  $0.14$ \\
	$D^+\to K_0^*$ & $5.6$ & $5.73\pm 0.35$~\cite{PDG} & $1.13$ 
	& $0.22$ \\
	$D^+\to \rho_0$ & $0.25$ & $0.25\pm 0.08$~\cite{PDG} & $1.11$ & $0.14$ \\
	$D^+\to \omega$ & $0.25$ & $0.17\pm0.06\pm0.01$~\cite{Blusk05} & $1.10$ & $0.14$\\
	$D_s\to \Phi$ & $2.4$ & $2.0\pm 0.5$~\cite{PDG} & $1.08$ & $0.21$ \\
	$D_s\to K_0^*$ & $0.22$ &  & $1.03$ & $0.13$ \\
\hline \hline
\end{tabular}
\caption{The branching ratios and partial decay width 
ratios for the $D\rightarrow V$ semileptonic decays. Comparison of our model 
fit with experiment as explained in the text.}
\label{table_results}
\end{center}
\end{table}

\section{Search for new physics in rare D decays}

At low-energies new physics is usually 
expected in the down-like quark sector. Numerous studies of new  physics 
effects  were performed in the $s \to d$, $b \to s (d)$, 
$\bar s d \leftrightarrow \bar d s$, $\bar b d \leftrightarrow \bar d b$ and 
$\bar b s \leftrightarrow \bar s b$ transitions.

However, searches for new physics in the up-like quark sector at low energies 
were not so  
attractive. 
Reasons are following: a)flavor changing neutral current processes 
at loop level in the standard model suffer from the  
 GIM cancellation leading to very small effects in the $c \to u$ 
transitions. 
 The GIM mechanism acts in many extensions of the standard model 
too, making contributions of new physics insignificant.  
b)   Most of the charm meson processes,  where 
$c \to u$  and $c \bar u \leftrightarrow \bar c u$ transitions might occur  
are  dominated by the standard model long-distance  contributions
 \cite{burdman1} - \cite{bigi0}.

On the experimental side there are many studies of rare charm meson decays. 
The first  observed rare $D$ meson decay was  the radiative
 weak decay 
$D \to \phi \gamma$. Its rate $BR(D \to \phi \gamma)= 2.6^{+0.7}_{-0.6} \times 10^{-5}$ has been 
measured by Belle collaboration 
\cite{Belle1}  and hopefully other
 radiative weak charm decays will be observed soon\cite{CLEO_pll}. 

\vspace{0.2cm}


In the standard model (SM) \cite{burdman1} the contribution coming from the penguin 
diagrams in 
$\rm c\to u\gamma$ transition gives 
branching ratio of order $10^{-18}$. 
The QCD corrected
effective Lagrangian \cite{greub} gives $\rm BR(c\to u\gamma)\simeq3\times10^{-8}$. 
A variety of models beyond SM  were
investigated and it was found that the gluino exchange diagrams
\cite{sasa} within general minimal supersymmetric SM (MSSM) might lead to  the
enhancement
\begin{equation}
\rm\frac{BR(c\to u\gamma)_{{MSSM}}}{BR(c\to u\gamma)_{{SM}}} 
\simeq10^2.
\label{1}
\end{equation}



Within SM the $c\to ul^+l^-$ amplitude is given by the $\gamma$ and $Z$ 
penguin
diagrams and $W$ box diagram. 
It is dominated by the light quark contributions in
the loop.  
The leading order rate for the inclusive $c\to u l^+l^-$ calculated within 
 SM \cite{prelovsek3}
was found to be suppressed by QCD corrections \cite{burdman2}. 
The inclusion of the renormalization group equations  
for the Wilson coefficients 
gave an additional significant 
suppression \cite{jure} leading to the rates  
$\Gamma(c\to ue^+e^-)/\Gamma_{D^0}=2.4\times 10^{-10}$ and
$\Gamma(c\to u\mu^+\mu^-)/\Gamma_{D^0}=0.5\times 10^{-10}$.   
These transitions are largely driven by virtual photon at low dilepton mass $m_{ll}$.

The leading 
contribution to $c\to ul^+l^-$ in general MSSM with conserved R parity 
comes from the one-loop diagram with 
gluino and squarks in the loop \cite{burdman2,prelovsek3,sasa}. 
It proceeds via virtual photon  
and significantly enhances the $c\to ul^+l^-$ 
spectrum at small dilepton mass $m_{ll}$. 
The authors of Ref. \cite{burdman2} have investigated supersymmetric 
(SUSY) extension of the SM with R parity breaking and they 
found that it can modify the rate. Using most resent CLEO \cite{CLEO_pll} 
results for the $D^+ \to \pi^+ \mu^+ \mu^-$ one can set the bound for the product of the 
relevant parameters entering 
the R parity violating $\tilde \lambda'_{22k} \tilde \lambda'_{21k} \simeq 0.001 $ 
(assuming that the 
mass of squark $M_{\tilde D_k} \simeq 100$ GeV). This bound gives the rates 
$BR_R(c\to ue^+e^-) \simeq1.6 \times 10^{-8}$ and  
$BR_R(c\to u \mu^+\mu^-) \simeq1.8 \times 10^{-8}$. 

Some  of models of new physics (NP) contain an extra up-like 
heavy quark 
inducing flavor changing neutral currents at tree 
level for the up-quark sector \cite{FP-LH,barger,lang,abel,higuchi}. 
The isospin component of the weak neutral current is given in \cite{FP-LH} as
\begin{equation}
J_{W^3}^\mu = \frac{1}{2} \bar U_L^m \gamma^\mu \Omega U_L^m -  
\frac{1}{2} \bar D_L^m \gamma^\mu  D_L^m
\label{e2}
\end{equation}
with $L=\frac{1}{2}(1- \gamma_5)$ and mass eigenstates 
$U_L^m= (u_L,c_L,t_L,T_L)^T$, $D_L^m=(d_L,s_L,b_L)^T$.
The neutral current for the down-like quarks is the same as in 
the SM, while there are tree-level flavor changing transitions 
between up-quarks if $\Omega \not =I$. The elements of $4\times 4$ matrix 
$\Omega$ can be constrained by CKM  unitarity violations 
currently allowed by experimental data. Even more stringent bound 
on $c u Z$ coupling $\Omega_{uc}$ comes from the present 
bound on $\Delta m $ in $ D^0 - \bar  D^0$ transition. 
It gives $|\Omega_{uc}| \leq 0.0004$ and we use the upper bound to 
determine the maximal effect on rare $D$ decays in what follows.
In this case the dilepton mass distribution of 
the $c \to u l^+l^-$  differential branching ratio can be enhanced 
by two orders of magnitude in comparison with SM (see Fig.~4). 

A particular version of the model with  
tree-level up-quark FCNC transitions is the Littlest Higgs model \cite{lee}.  
 In this case the  magnitude of the relevant $c \to u Z$ coupling  
$\Omega_{cu} =|V_{ub}||V_{cb}|v^2/f^2 \leq 10^{-5}$ 
is even further 
constrained via the scale $f\geq {\cal O}(1~{\rm TeV})$ 
by the precision electro-weak data. The smallness of $\Omega_{uc}$
 implies that the effect of this particular 
model on $c\to ul^+l^-$ decay and relevant rare $D$ decays is insignificant
 \cite{FP-LH}. 



The study of exclusive D meson rare decay modes is very difficult due to the 
dominance of the long distance effects \cite{burdman1} - \cite{prelovsek2}.
The inclusive $c \to u l^+ l^-$ can be tested in the rare decays 
$D \to \mu^+ \mu^-$, 
$D \to P (V) l^+ l^-$ \cite{burdman2,prelovsek3,burdman3}.

The branching ratio for the rare decay $D\to \mu^+ \mu^-$ 
is very small in the SM. 
The detailed treatment of this decay rate \cite{burdman2} 
gives $Br(D \to \mu^+ \mu^-) \simeq 3\times 10^{-13}$ \cite{burdman2}. This decay rate 
can be enhanced within a study which considers 
 SUSY with R parity breaking effects \cite{burdman2,bigi0}. 
Using the bound $\tilde \lambda'_{22k} \tilde \lambda'_{21k} \simeq 0.001 $ 
one obtains the limit $Br(D \to \mu^+ \mu^-)_R\simeq 4\times 10^{-7}$. 

The $D \to P (V) l^+ l^-$ decays offer another possibility to study the $c \to u l^+ l^-$ transition in charm sector. 
The most appropriate decay modes for the experimental searches 
are $D^+\to \pi^+ l^+l^-$ and  
$D^0\to \rho^0e^+e^-$.  In the following we present the possible maximal 
effect on these decays coming from a general class of models with tree level 
$cuZ$ coupling at its upper bound $|\Omega_{uc}|=0.0004$. 
We already pointed out that in the Littlest Higgs model, 
which is a particular version of these models, 
 the coupling $\Omega_{uc}$ is constrained to be 
smaller and the effects on rare $D$ decays are insignificant \cite{FP-LH}.

The    calculations of the long distance contributions in the  decays 
$D^+\to \pi^+l^+l^-$ and $D^0 \to \rho^0 l^+ l^-$ are presented 
in Refs. \cite{FP-LH,prelovsek2,prelovsek3}.
The contributions of the 
intermediate vector resonances $V_0=\rho^0,\omega,\phi$ with $V_0\to l^+l^-$ 
constitute 
an important long-distance contribution to the hadronic decay, 
which may shadow interesting 
short-distance contribution induced by $c\to ul^+l^-$ transition. 

Our determination of short and long distance contributions to 
$D^+\to \pi^+l^+l^-$ takes advantage of the available 
experimental data \cite{FP-LH}. 
This is a fortunate circumstance for this particular decay since the 
analogous experimental input is not available for determination of  
the other $D\to X l^+l^-$ rates in a similar way. The rate 
resulting from the amplitudes (14) and (19) of \cite{FP-LH} with   
$|\Omega_{uc}|=0.0004$ are given  in Fig.~5 and Table~4.

We are unable to determine the  amplitude of 
the long-distance contribution to $D^0\to \rho^0 V_0\to \rho^0 l^+l^-$ 
using the measured rates for 
$D^0\to \rho^0 V_0$ since only  the rate of $D^0\to \rho^0 \phi$ is known 
experimentally. We are forced to use a model \cite{prelovsek2}, 
developed to describe all $D\to Vl^+l^-$ and $D\to V\gamma$ decays, 
and the resulting rates are presented in Fig.~6 and Table~4.

\begin{figure}[!t]
\begin{center}
\epsfig{file=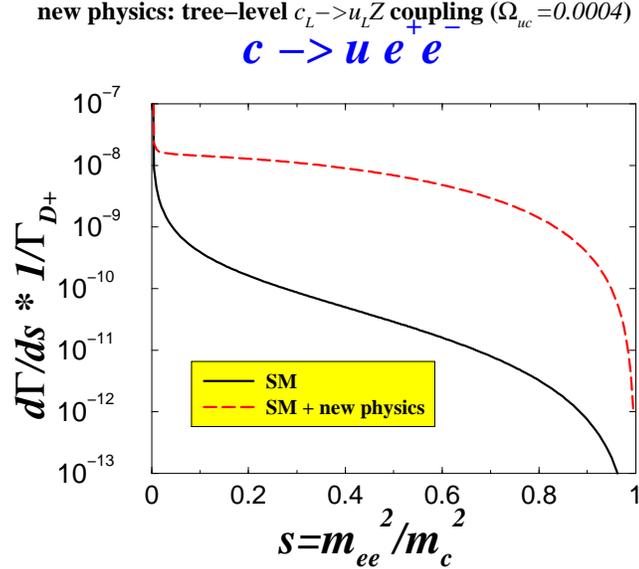,width=3.6in}
\end{center}
\caption{    The dilepton mass distribution 
$dBr/dm_{ee}^2$ for the inclusive decay 
$c\to u l^+l^-$   
as a function of the  dilepton mass square $m_{ee}^2=(p_++p_-)^2$.
}
\end{figure}

Therefore, the total rates for $D \to X l^+ l^-$ are dominated by the 
long distance resonant contributions at dilepton mass 
$m_{ll}=m_\rho,~m_\omega,~m_\phi$ 
and even the largest contributions from new physics are not expected to 
affect the total rate significantly \cite{burdman2,prelovsek3}. 
New physics could only modify the SM 
differential spectrum at low $m_{ll}$ below 
$\rho$ or spectrum at high $m_{ll}$ above $\phi$. 
 In the case of $D\to\pi l^+l^-$ differential decay 
distribution there is a broad  
 region at high $m_{ll}$ (see Fig.~5), 
 which presents a unique possibility to 
study $c\to ul^+l^-$ transition \cite{prelovsek3,FP-LH}.

\begin{table*}[!b]
\begin{center}
{\begin{tabular}{@{}ccccccc@{}}
\hline \hline
 {\bf Br} & \multicolumn{2}{c|}{short distance } & total rate $\simeq$ & experiment\\
 & \multicolumn{2}{c|}{contribution only } & long distance contr. & \\  
\hline 
 & SM & SM + NP &    &   \\
\hline
$D^+\to \pi^+ e^+e^-$ & $6\times 10^{-12}$ & $8\times 10^{-9}$ & $1.9\times 10^{-6}$ & $<7.4\times 10^{-6}$\\
$D^+\to \pi^+ \mu^+\mu^-$ &  $6\times 10^{-12}$ & $8\times 10^{-9}$ & 
$1.9\times 10^{-6}$ & $<8.8\times 10^{-6}$\\
\hline
$D^0\to \rho^0 e^+e^-$ & negligible &$5\times 10^{-10}$ & $1.6\times 10^{-7}$
&$<1.0\times 10^{-4}$\\
$D^0\to \rho^0 \mu^+\mu^-$ & negligible &$5\times 10^{-10}$ & $1.5\times 10^{-7}$ & $<2.2\times 10^{-5}$\\
\hline \hline
\end{tabular}}
\caption{Branching ratios for the decays in which $c\to ul^+l^-$ transition 
can be probed.} 
\label{tab1}
\end{center}
\end{table*}

The non-zero forward-backward asymmetry in $D\to \rho l^+l^-$ decay  arises 
only when $C_{10}\not =0$ (assuming $m_l\to 0$).  
The enhancement of the $C_{10}$ in the NP  models \cite{FP-LH} 
 is due to the tree-level $\bar u_L\gamma_\mu c_LZ^\mu$ coupling
 and leads to  nonzero asymmetry $A_{FB}(m_{ll}^2)$ shown in 
Fig.~7.   
The forward-backward asymmetry for  $D^0 \to \rho^0 l^+ l^-$ vanishes in 
SM ($C_{10}\simeq 0$), while it is reaching ${\cal O}(10^{-2})$  
in NP model with the extra up-like quark as shown in Fig.~7. 
Such asymmetry is still small and  difficult to be seen in 
the present or planned experiments given that the rate itself is already 
small.

\begin{figure}[!t]
\begin{center}
\epsfig{file=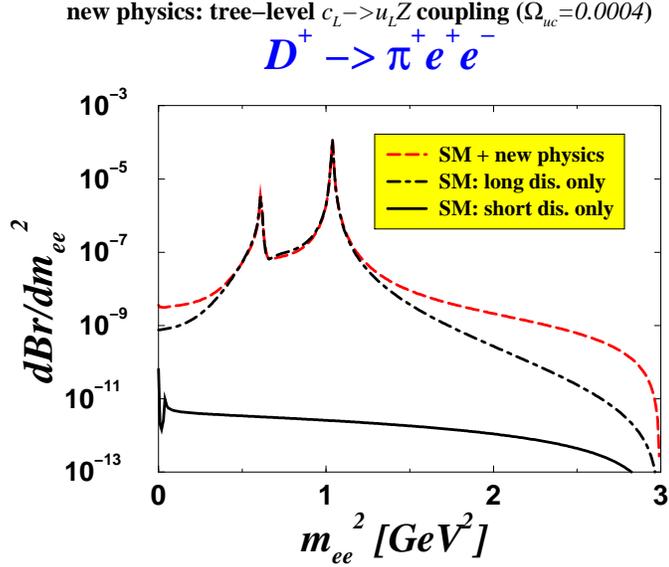,width=3.6in}
\end{center}
\caption{The dilepton mass distribution 
 $dBr/dm_{ee}^2$ for $D^+\to \pi^+e^+e^-$.   }
\label{fig2}
\end{figure}
\begin{figure}[!t]
\begin{center}
\epsfig{file=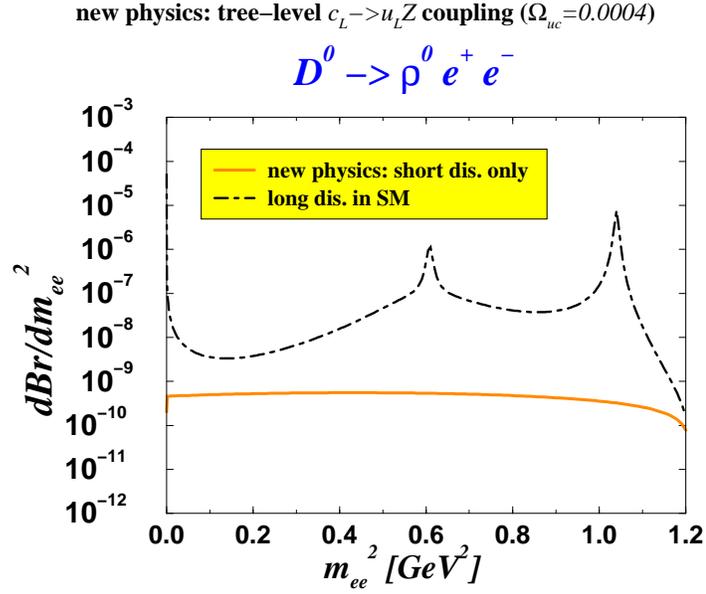,width=3.6in}
\end{center}
\caption{ 
The dilepton mass distribution for $D^0\to \rho^0e^+e^-$.} 
\end{figure} 
\begin{figure}[!t]
\begin{center}
\epsfig{file=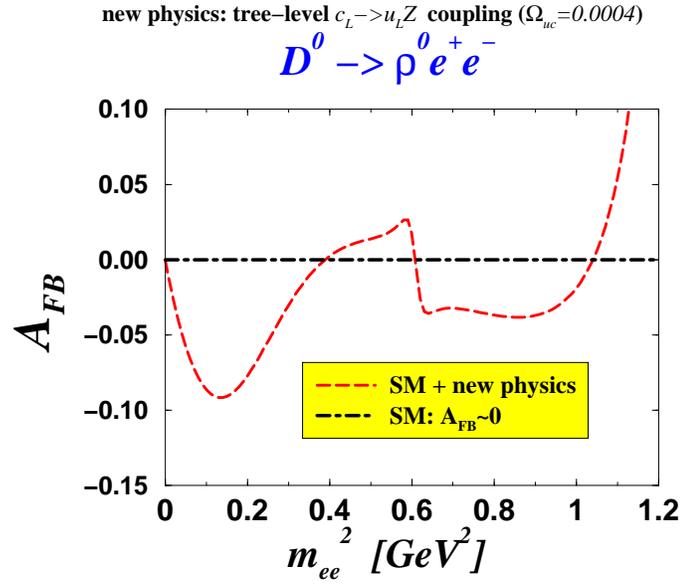,width=3.6in}
\end{center}
\caption{  
 The forward-backward asymmetry for $D^0\to \rho^0e^+e^-$.}
\end{figure}



We have investigated impact   of the tree-level flavor changing 
neutral transition $c \to u Z$ on the rare $D$ meson decay observables. 
However, the most suitable 
$D^+ \to \pi^+ l^+l^-$ and $D^0 \to \rho^0 l^+ l^-$ decays are found to be 
dominated by the SM long distance contributions. Only small enhancement 
of the differential mass distribution can be seen in the case of $D^+ \to \pi^+ l^+ l^-$ decay at high dilepton mass and  tiny forward backward asymmetry can be induced by new physics in $D^0 \to \rho^0 l^+ l^-$ decay.

We conclude that the NP scenarios which contain an extra singlet heavy up-like quark,  have rather small effects on the 
charm meson observables. 





\begin{thebibliography}{99}


\bibitem{SF-JK-ch} S. Fajfer and J. Kamenik, Phys. Rev. D {\bf 74}, 074023 (2006).



\bibitem{Stewart} 
I.W. Stewart, Nucl. Phys. B {\bf529}, 62 (1998).


\bibitem{mehen} T. Mehen and R. Springer, Phys. Rev. D {\bf 72}, 034006 (2005).


\bibitem{Abada} A. Abada et al., J. High Enegry Phys. 02, 016,  (2004).

\bibitem{McNeile} C. McNeile, C. Michael, and G. Thompson (UKQCD), 
Phys. Rev D {\bf 70}, 054501 (2004). 

\bibitem{PDG}
W.-M. Yao et al., Journal of Physics G 33, 1  (2006). 

\bibitem{Huang04} Huang, et~al., Phys. Rev. Lett. {\bf 94}, 011802 (2005).

\bibitem{Link04}
J.~M. Link, et~al., Phys. Lett. B {\bf 607}, 233 (2005).

\bibitem{Becirevic99}
D.~Becirevic, and A.~B. Kaidalov, Phys. Lett. B {\bf 478}, 417
  (2000). 

\bibitem{Aubert03}
B.~Aubert, et~al., Phys. Rev. Lett. {\bf 90}, 242001 (2003).

\bibitem{Vaandering04}
E.~W. Vaandering  (2004), hep-ex/0406044.

\bibitem{Besson03}
D.~Besson, et~al., AIP Conf. Proc. {\bf 698}, 497 (2004).

\bibitem{Evdokimov04}
A.~V. Evdokimov, et~al., Phys. Rev. Lett. {\bf 93}, 242001 (2004).
\bibitem{Fajfer04}
S.~Fajfer, and J.~Kamenik, Phys. Rev. D {\bf 71}, 014020
  (2005). 

\bibitem{Fajfer05}
S.~Fajfer, and J.~Kamenik, Phys. Rev. {\bf 72}, 034029
  (2005). 

\bibitem{Fajfer06}
S.~Fajfer, and J.~Kamenik, Phys. Rev. {\bf 72}, 057503 
  (2006).


\bibitem{Marshak69}
R.~E. Marshak, Riazuddin, and C.~P. Ryan, {\it Theory of Weak Interactions in
  Particle Physics}, vol. XXIV of {\it Interscience Monographs and Texts in
  Physics and Astronomy}, Wiley-Interscience, New York, 1969.

\bibitem{Wirbel85}
M.~Wirbel, B.~Stech, and M.~Bauer, Z. Phys. C {\bf 29}, 637 (1985).

\bibitem{Charles98}
J.~Charles, A.~Le~Yaouanc, L.~Oliver, O.~Pene, and J.~C. Raynal, Phys.
  Rev. D {\bf 60}, 014001 (1999). 

\bibitem{Isgur90}
N.~Isgur, and M.~B. Wise, Phys. Rev. D {\bf 42}, 2388 (1990).

\bibitem{Hill05}
R.~J. Hill  (2005), hep-ph/0505129.

\bibitem{Ebert01}
D.~Ebert, R.~N. Faustov, and V.~O. Galkin, Phys. Rev. D {\bf 64},
  094022 (2001). 

\bibitem{Burdman00}
G.~Burdman, and G.~Hiller, Phys. Rev. D {\bf 63}, 113008 (2001).

\bibitem{Hill04}
R.~J. Hill  (2004), hep-ph/0411073.


\bibitem{Bajc95}
B.~Bajc, S.~Fajfer, and R.~J. Oakes, Phys. Rev. D {\bf 53}, 4957
  (1996). 

\bibitem{Ball91}
P.~Ball, V.~M. Braun, and H.~G. Dosch, Phys. Rev. {\bf D44},
  3567--3581 (1991).

\bibitem{Melikhov00}
D.~Melikhov, and B.~Stech, Phys. Rev. D {\bf 62}, 014006 (2000).

\bibitem{Aubin04}
C.~Aubin, et~al.  (2004), hep-ph/0408306.

\bibitem{Abada02}
A.~Abada, et~al., Nucl. Phys. Proc. Suppl. B {\bf 119}, 625
  (2003). 

\bibitem{Ball93}
P.~Ball, Phys. Rev. D  {\bf 48}, 3190 (1993).

\bibitem{Blusk05}
S.~Blusk  (2005), hep-ex/0505035.
\bibitem{FP-LH} S. Fajfer and Sasa Prelovsek, 
Phys. Rev.D {\bf 73}, 054026 (2006). 

\bibitem{burdman1} G. Burdman, E. Golowich, J. Hewett and 
S. Pakvasa, Phys. Rev. D {\bf 52}, 6383 (1995).

\bibitem{burdman2} G. Burdman, E. Golowich, J. Hewett and 
S. Pakvasa, Phys. Rev. D {\bf 66}, 014009 (2002).

\bibitem{burdman3} G. Burdman and I. Shipsey, Ann. Rev. Nucl. 
Part. Sci. {\bf 53} 431 (2003).

\bibitem{pakvasa} S. Pakvasa, Nucl. Phys.B Proc. Suppl. {\bf 142},
115 (2005).

Durham, England, Apr 2003, hep-ph/030626.  


\bibitem{prelovsek1} S. Fajfer, S. Prelovsek, P. Singer,
Eur. Phys. J.C {\bf 6}, 471 (1999). 

\bibitem{prelovsek2} S. Fajfer, S. Prelovsek, P. Singer, 
Phys. Rev. D {\bf 58}, 094038 (1998). 

\bibitem{prelovsek3} S. Fajfer, S. Prelovsek, P. Singer, 
Phys. Rev.D {\bf 64}, 114009 (2001). 

\bibitem{jure} S. Fajfer, P. Singer, J. Zupan, 
Eur. Phys.J.C {\bf 27}, 201 (2003).

\bibitem{bigi0} S. Bianco, F.L. Fabbri, D. Benson and I. Bigi,
Riv. Nuovo 
Cim. 26 N {\bf 7}, 1 {2003}. 

\bibitem{Belle1} O. Tajima  {\it et al.}, BELLE Collaboration,
Phys. Rev. Lett. {\bf 92},  101803 (2004). 

\bibitem{CLEO_pll} Q. He {\it et al.}, CLEO Collaboration, 
Phys. Rev. Lett.{\bf 95}, 221802 (2005).
\bibitem{greub} C. Greub, T. Hurth, M. Misiak and D. Wyler, 
Phys. Lett. B {\bf 382}, 415 (1996);  
Q. Ho Kim and X.Y. Pham, Phys. Rev. D {\bf 61}, 013008 (2000).
\bibitem{sasa} S. Prelovsek and D. Wyler, Phys. Lett.B {\bf 500}, 304 
(2001).
\
\bibitem{barger} V. Barger, M. S. Berger and R.J. N. Phillips, 
Phys. Rev. D {\bf 52}, 1663 (1995).

\bibitem{lang} P. Langacker and D. London, Phys. Rev. D {\bf 38}, 886 (1988).

\bibitem{aguila} F. del Aguila and J. Santiago, JHEP {\bf 03}, 010 (2002).


\bibitem{abel} S.A. Abel, J. Santiago, M. Masip, 
JHEP {\bf 04}, 057 (2003).
\bibitem{higuchi} K. Higuchi and K. Yamamoto, Phys. Rev. D {\bf 62}, 03005
 (2000). 
\bibitem{lee} Lae Yong Lee, JHEP {\bf 0412}, 065 (2004).












\end{thebibliography}
\end{document}